\documentclass{aa}  
\usepackage{xcolor}

\begin{document}

\title{Spectropolarimetry of the changing-look active galactic nucleus NGC~1566 and its potential link to supermassive black hole binaries}
\titlerunning{Spectropolarimetry of NGC~1566}%
   \author{F.~Marin\inst{1}          
          \and
          V.~H.~Sasse\inst{2}    
          \and
          J.~Biedermann\inst{1} 
          \and
          D.~Hutsem\'ekers\inst{3}  
          \and
          R.~C.~Fernandes\inst{2}    
          \and
          D.~Porquet\inst{4}   
          \and
          V.~Oknyansky\inst{5,6} 
          }

   \institute{Universit\'e de Strasbourg, CNRS, Observatoire Astronomique de Strasbourg, UMR 7550, 11 rue de l'universit\'e, 67000 Strasbourg, France\\
             \email{frederic.marin@astro.unistra.fr}
             \and            
                 Universidade Federal de Santa Catarina, Campus Universitário Reitor João David Ferreira Lima, Florianópolis 88040-900, Brazil
             \and            
                 Institut d’Astrophysique et de G\'eophysique, Universit\'e de Li\`ege, All\'ee du 6 Ao\^ut 19c, B5c, 4000 Li\`ege, Belgium
             \and            
                 Aix Marseille Univ, CNRS, CNES, LAM, Marseille, France 
             \and            
                 Sternberg Astronomical Institute, M.V. Lomonosov Moscow State University, 119234, Moscow, Universitetsky pr-t, 13, Russia 
             \and            
                 Department of Physics, Faculty of Natural Sciences, University of Haifa, Haifa 3498838, Israel 
             }

   \date{Received January 12, 2026; accepted March 25, 2026}


\abstract
{The active galactic nucleus (AGN) NGC~1566 is known to present dramatic and regular spectral shape changes, associated with the appearance and disappearance of broad emission lines.}
{The underlying mechanism responsible for such changes is yet to be identified, but occultation, eccentric accretion disks, turbulent disk-dominated broad line regions (BLRs) or binary supermassive black holes have been hypothesized.}
{Because the scenarios used to explain the variable spectral shapes of NGC~1566 each have a specific geometric configuration, we used the Very Large Telescope (VLT) FOcal Reducer/low dispersion Spectrograph 2 (FORS2) instrument to obtain nine 3500-10\,000~\AA\, polarized spectra of the source between August 2 and September 21, 2025.}
{We caught the AGN in a type-2 state, i.e., without any broad component in total nor polarized fluxes. Its low (0.22-0.24\%) and wavelength-independent polarization degree (and angle) above 4000~\AA\, argues against occultation of the BLR and is consistent with a significant weakening or disappearance of the BLR. The polarized spectrum reveals a strong rise of polarization in the blue band, likely echoing the 2018 outburst of the AGN. The temporal variability of the total flux continuum but the steadiness of the line profiles demonstrate that the object is viewed close to pole-on, irrespective of its spectral type at the time of observation. Relative to archival data, NGC 1566 shows significant variability in polarization degree, angle, and wavelength dependence, transitioning from gray to chromatic polarization. Even more surprisingly, NGC~1566 behaves opposite to the basic predictions of the unified model: its polarization angle is perpendicular to the AGN polar axis and its polarization degree is higher when in a brighter, type-1 phase.}
{The results reported above contradict occultation and binary supermassive black hole hypotheses. They support, rather, accretion-driven photoionization and/or structural changes in the internal accretion flow and the BLR, but their geometry can only be probed by new spectropolarimetric campaigns during a type-1 phase of NGC~1566.}


\maketitle
\nolinenumbers 
\section{Introduction}
\label{Introduction}%

NGC~1566 is an odd active galactic nucleus (AGN). Situated in the nearby Universe ($z$ = 0.005017), at coordinates RA = 04h20m00.3940s, Dec = -54d56m16.598s (65.001642$^\circ$, -54.937944$^\circ$), it is the brightest member of the southern Dorado Group. NGC~1566's host is classified as an SABbc galaxy \citep{Vaucouleurs1973} with no nearby companions and its central supermassive black hole (SMBH) mass was estimated using several methods, converging toward a relatively light measure of 5 $\times$ 10$^6$~M$_\odot$ \citep{Ochmann2024}. Added to its weak bolometric luminosity (1.4-4 $\times$ 10$^{41}$~erg~s$^{-1}$, \citealt{Slater2019}) and accretion rate (2.5-7.1 $\times$ 10$^{-5}$~M$_\odot$~y$^{-1}$, ibid.), NGC~1566 is classified as a low-luminosity AGN. It also belongs to the Seyfert class, according to the absence of any significant radio flux nor prominent radio structure, as imaged by \citet{Morganti1999} using the Australia Telescope Compact Array (ATCA) at 3.5~cm (8.6~GHz), with no signs of jets or lobes and a rather centro-symmetrical radio emission pattern. Since its first known spectroscopic observation, this AGN has been classified as a type-1 object \citep{Vaucouleurs1961}, meaning that its optical spectrum is composed of a power-law continuum and bright, narrow (full width at half-maximum -- FWHM -- $\le$ 1000~km~s$^{-1}$), and broad (FWHM $\ge$ 2000~km~s$^{-1}$) emission lines, the Balmer series lines being the widest of those observed \citep{Shobbrook1966}.

However, it was swiftly detected that NGC~1566's spectrum behaved differently to regular Seyfert galaxies. \citet{Pastoriza1970} reported a significant weakening of the broad H$\beta$ emission in 1969, yet still a broad H$\alpha$ profile blended with [N II] forbidden emission. An intensive monitoring campaign, both in photometry and spectroscopy, led by \citet{Alloin1985,Alloin1986} between 1970 and 1985, showed that the Balmer series lines vary significantly in concordance with the source of ionizing flux, as the region responsible for the emission of the broad lines (the BLR) is rather compact (less than 20 light-days, \citealt{Oknyanskij2001}). Over a period of 15 years, the active nucleus has presented four activity cycles of about 1300 days each, alternating between short (weeks-long) yet bright periods with year-long, low-luminosity phases. For example, in the 1981–1985 cycle, several bright and fast outbursts were detected. Such outbursts rose in flux in about 20 days then had an exponential decrease lasting about 400 days. During those outbursts, the AGN continuum increased in flux and became bluer, matched with more intense Balmer emission. \citet{Alloin1986} demonstrated, based on timing and energetic arguments, that supernova explosions could not be at the origin of the activity in NGC~1566. The authors rather advocated for an accretion-related origin. Indeed, thanks to high-frequency spectroscopic monitoring between 1980 and 1982, \citet{Alloin1985} discovered that NGC~1566 was able to transition from a type-1 to a type-2 classification (i.e., without any broad emission features in total flux) in less than four months. Such a transition is accompanied by a rise in non-stellar continuum flux and proportional flux variations in the broad Balmer emission lines, in accordance with photoionization models, once again supporting the spectral transitions \citep{Abramowicz1986} having an accretion-driven origin. However, an intriguing -- and yet to be explained -- feature of the outbursts is that the broad emission lines appear to vary before the optical continuum (i.e., the exact opposite of photoionization models), suggesting that the optical and ionizing non-stellar flux may not be directly linked, but instead coupled via slower secondary processes. In addition, the broad line profiles are asymmetric but constant in time and velocity, something unusual for Seyfert galaxies, indicating that the BLR clouds are rather radiation-bounded (i.e., the BLR clouds are thick enough to absorb all the ionizing photons that hit them) than matter-bounded (when the BLR clouds are too thin to absorb all ionizing photons). Finally, exploring the non-varying profile and strength of the narrow forbidden lines, \citet{Alloin1985} concluded that NGC~1566 probably spends much more time in the low-luminosity, type-2 state than in the bright, type-1 one.

This variability in time is actually supported by X-ray to infrared (IR) observing campaigns \citep{Baribaud1992,Elvis1990,Oknyansky2019}. In the radio band, the flux of the AGN is relatively weak (8~mJy, \citealt{Morganti1999}) and there is no detected jet or any synchrotron-related structure. Looking for anisotropies in the AGN flux distribution, Hubble Space Telescope (HST) high-resolution narrow-band [O~III] images revealed a one-sided ionization cone toward the southeast \citep{Schmitt1996}. The shape of the emission source was, at that time, not spatially resolved enough to define a preferred ejection position angle. With better instrumentation, namely integral-field spectroscopy data, \citet{daSilva2017} also examined this [O III] outflow embedded in a ionized gas and reported a mean position angle of 122$^\circ$ $\pm$ 16$^\circ$. They also noted that this outflow is almost perpendicular to the line of nodes (position angle 13$^\circ$ $\pm$ 2$^\circ$) that is thought to be associated with the molecular disk seen in H$_2$. Such a measurement agrees with Einstein High Resolution Imager (HRI) X-ray maps, which revealed that the high-energy emission is very asymmetric in NGC~1566, extending along a position angle of $\sim$ 315$^\circ$ on a 2.5"-10"-scale \citep{Elvis1990}. This makes it the best estimate of the nuclear symmetry axis of NGC~1566 so far.

As of today, the source continues to show outbursts and varying optical spectral classification, visiting almost all known classifications (1.2, 1.5, 1.8, 1.9, see \citealt{Oknyansky2019}), which led to NGC~1566 being included in the recurrent “changing-look” AGN category. The source flux is regularly increasing and decreasing by more than one magnitude \citep{Oknyansky2019}, driving the appearance and disappearance of the broad emission lines, with significant strengthening of the ultraviolet (UV) Balmer continuum and high-ionization coronal lines, such as [Fe~X]~$\lambda$6374 \citep{Oknyansky2019,Oknyansky2020}. Even more than that, broad double-peaked emission line profiles of O~I~$\lambda$8446 and the Ca II~$\lambda\lambda$8498, 8542, 8662, triplet, never ever reported in the AGN literature before, suddenly and recently appeared in the spectrum of NGC~1566 \citep{Ochmann2024}. In the type-1 phase reported by the authors, all broad lines showed a clear redward asymmetry with respect to their central wavelength. To explain such variations, \citet{Ochmann2024} suggest that an eccentric accretion disk associated with a turbulent, disk-dominated BLR lies at the heart of NGC~1566. If correct, one would expect strong polarization signatures, both in the continuum and in the broad emission lines, since asymmetry enhances polarization and flared BLR structures can achieve polarization degrees of up to several percent, depending on their half-opening angle and optical depth \citep{Goosmann2007,Marin2012}. Yet, the polarization from this source was only measured twice in the past (see Sect.~\ref{Analysis:old_data}). 

The key of the changing-look enigma could be linked with the very recent discovery of a gravitational lensing outburst in NGC~1566, which revealed systematical velocity shifts of the broad emission line profiles together with a steep symmetric light curve, very different with respect to normal random variations in AGNs \citep{Kollatschny2024}. According to the authors, those features are symptomatic of a sub-parsec SMBH binary in the nuclear region of NGC~1566. The link between changing-look AGNs and gravitationally bound SMBHs is particularly tempting, since the spectral type variations and the flux brightening (or dimming) are most naturally explained by a black hole orbiting another, perturbing the central accretion structure and the BLR \citep{Wang2020}. Here again, the presence of a binary component would result in prominent, time-dependent geometrical asymmetries in the core of NGC~1566, and hence variable polarizations of the continuum and broad emission lines \citep{Savic2019}. Spectropolarimetry, from the UV to the NIR bands, is thus the best-suited tool to confirm or reject 1) an occultation of the BLR, 2) an eccentric accretion disk, 3) a turbulent, disk-dominated BLR, and 4) a binary SMBH, all at once. For this reason, we applied for Very Large Telescope (VLT) FOcal Reducer/low dispersion Spectrograph 2 (FORS2) spectropolarimetric time and present the results of our observing campaign in this paper. 

\section{Observation and data reduction}
\label{Observation}%

\subsection{VLT/FORS2 observation}
\label{Observation:VLT}%

\begin{table}
    \caption{Observation log.}%
    \centering%
    \begin{tabular}{lcccc}
        \textbf{Date} & \textbf{Grism} & \textbf{Time} & \textbf{Airmass} & \textbf{Seeing}\\
        \textbf{2025} & & \textbf{(s)} & & \textbf{$\arcsec$}\\
        \hline
        Aug 3  & 300V       & 2 $\times$ 1200  & 1.4-1.3 & 0.8\\
        Aug 21 & 300V       & 2 $\times$ 1200  & 1.3-1.2 & 1.1 \\
        Aug 22 & 300I+OG590 & 2 $\times$ 1200  & 1.3-1.2 & 1.0 \\
        Aug 23 & 300V       & 2 $\times$ 1200  & 1.4-1.3 & 1.1 \\
        Sep 18 & 300I+OG590 & 2 $\times$ 1200  & 1.3-1.2 & 0.5 \\
        Sep 20 & 300V       & 2 $\times$ 1200  & 1.4-1.3 & 0.6 \\
        Sep 20 & 300I+OG590 & 2 $\times$ 1200  & 1.2     & 0.9 \\
        Sep 21 & 300V       &            1200  & 1.2     & 0.7 \\
        Sep 21 & 300I+OG590 & 3 $\times$ 1200  & 1.2     & 0.7 \\
        \hline
    \end{tabular}%
    \tablefoot{We indicate the observation date, the grism + filter, the integration time (1200s corresponding to 4 $\times$ 300s for each HWP position), the airmass at the beginning and end of the observation, and the average seeing in arcseconds.}%
    \label{Tab:Obs}%
\end{table}

Spectropolarimetric observations were obtained using the VLT at the European Southern Observatory (ESO), which was equipped with the FORS2 instrument. Linear spectropolarimetry was performed by inserting a Wollaston prism into the beam to split the incoming light rays into two orthogonally polarized beams, separated by 22" on the charge-coupled device (CCD) detector. Sequences of four frames were obtained with the half-wave plate rotated at four different position angles (0$^\circ$, 22.5$^\circ$, 45$^\circ$, and 67.5$^\circ$) to derive the normalized Stokes parameters, $u$ and $q$. This combination allowed us to remove most of the instrumental polarization. The sequences were repeated twice per night, except on September 21. This provided a total of nine sequences at epochs ranging from August~3 to September~21.  At all epochs the sky was clear (i.e., less than 10\% of the sky above 30$^\circ$ elevation was covered in clouds, and transparency variations were under 10~\%). The observational circumstances are summarized in Table~\ref{Tab:Obs}.

We acquired the spectra with grisms 300V (3300–6600~\AA) and 300I+OG590 (6000–11000~\AA). The slit width was 1" on the sky, providing an average resolving power of R $\approx$ 440 (with grism 300V) and R $\approx$ 660 (with grism 300I). The slit was positioned in the north-south direction. CCD pixels were binned 2$\times$2, corresponding to a spatial scale of 0.25" per binned pixel.  We also observed polarized (BD-12~5133, BD-14~4922, and Hiltner 652) and unpolarized (WD1615-154, WD1620-391, and WD2149+021) standard stars \citep{Fossati2007}. The estimated instrumental polarization from the unpolarized standard stars is lower than 0.1\%. This is compatible with the instrumental polarization accurately measured by ESO\footnote{www.eso.org/sci/facilities/paranal/instruments/fors/inst/pola.html}, $P_{\rm inst}$ = 0.03\%.

\subsection{Data reduction}
\label{Observation:Reduction}

First, raw frames were processed to remove cosmic-ray hits using the Python implementation of the \textsc{lacosmic} package \citep{Dokkum2001,Dokkum2012}. Then, the ESO FORS2 pipeline \citep{Izzo2019} was used to obtain two-dimensional spectra that were flat-fielded, rectified, and calibrated in wavelength. One-dimensional spectra were extracted using apertures centered on the target with lengths of 11 pixels (2.75"), 21 pixels (5.25"), and 41 pixels (10.25"). The sky spectrum was estimated from adjacent multi object spectroscopy (MOS) strips and subtracted from the target spectrum. The normalized Stokes parameters, $u$ and $q$, were then computed from the ordinary and extraordinary spectra according to the procedure described in the FORS2 manual\footnote{https://www.eso.org/sci.html}, and corrected for the half-wave plate chromatic dependence. The total flux was corrected for the atmospheric extinction and calibrated using a master response curve. Uncertainties were estimated by propagating the photon and readout noises. Finally, the polarization degree, $P$, and polarization angle, $\theta$, were computed using the standard formulae, $P$ = $\sqrt{q^{2}+u^{2}}$ and 
$\theta$ = $\frac{1}{2}\arctan\,(\frac{u}{q})$ + $\Delta$, where $\Delta = 0^\circ$ for $u > 0$ and $q > 0$; $90^\circ$ for $q < 0$; and $180^\circ$ for $u < 0$ and $q > 0$, according to the usual north-south convention (0$^\circ$ in the north, 90$^\circ$ in the east). The polarization data were extracted independently for each of the nine series of four frames obtained with a grism.

\subsection{Averaged versus median values}
\label{Observation:AvgMed}

Because we could not reach a statistically significant signal-to-noise ratio (S/N) in polarization per resolution element without gathering all observations at once, we had to merge the nine exposures before computing the total flux, polarization degree, and angle. Since $P$ and $\theta$ are nonlinear functions of $q$ and $u$, it makes the combination method important. If we were to combine exposures by averaging or taking the median of the final polarization products ($P$, $\theta$), we would bias the result and degrade the S/N, as the noise distribution of polarization is non-Gaussian (it follows a Ricean distribution). The statistically correct approach is to combine the Stokes parameters themselves, using either the weighted average or the median of I, $q$, and $u$, and then compute $P$ and $\theta$. 

Both methods have pros and cons. For example, weighted averages are linear and minimize variance if noise is Gaussian, so they preserve flux and polarization information correctly, while medians do not minimize variance for Gaussian noise and the effective S/N gain is worse than for averages. However, averages are sensitive to outliers (cosmic rays, sky flare, tracking error, variable instrumental effects, etc.) that can bias the mean, while medians are more robust to outliers. In practice, both are often used interchangeably, but because NGC~1566 is known for its spectral variations we wanted to check that there was no risk in using either method.

\begin{figure}
    \centering%
    \includegraphics[width=\linewidth]{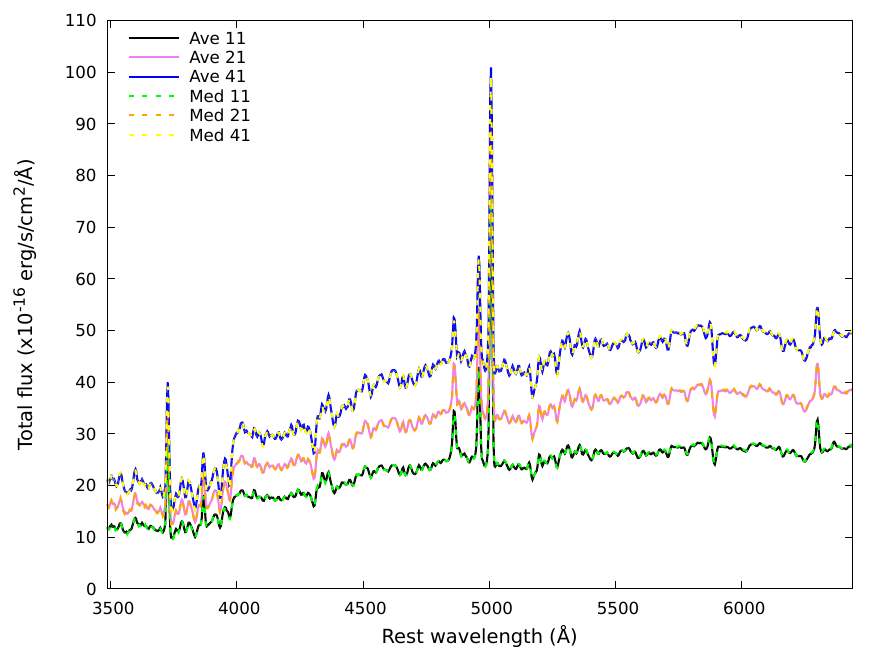}
    \includegraphics[width=\linewidth]{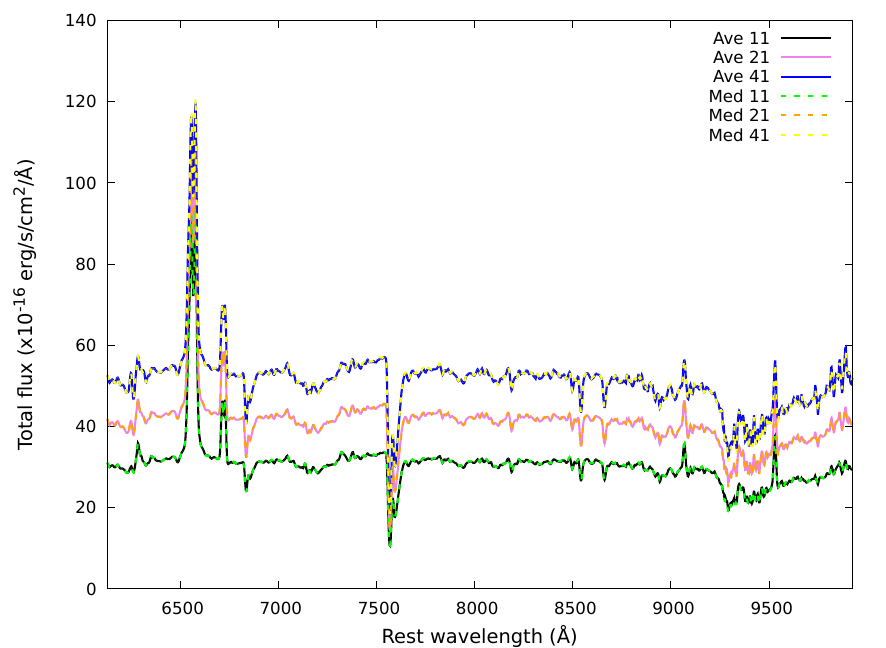}
    \caption{Total-flux spectra obtained using different slit lengths and different combination methods (average and median), applied to the merged dataset of NGC~1566. Top panel: Fluxes for the FORS2 300V grism. Bottom: Same but for the 300I grim (+ OG590 filter).}%
    \label{Fig:AvgMed_Flux}%
\end{figure}

In Fig.~\ref{Fig:AvgMed_Flux}, we present the averaged and median values of the total flux spectra in each grism (300V top, 300I bottom), for three different slit lengths: 11 pixels, 21 pixels, and 41 pixels. As we can see, both approaches give very similar results. The difference between the median and average method is less than 0.46~\% for the 300V grism and less than 0.32~\% for the 300I grism, averaged over the whole waveband. It is slightly higher than the averaged observational error (0.1\%). We therefore conclude that, for the total flux spectrum, both the median and the averaged combination produce nearly identical results, with differences well below the percent level and only slightly above the statistical noise. Also accounting for the fact that sigma(median) $\sim$ 1.25 $\times$ sigma(mean), either method is acceptable for combining the intensity spectra.

\begin{figure}
    \centering%
    \includegraphics[width=\linewidth,trim={0cm 0.5cm 0cm 0cm}]{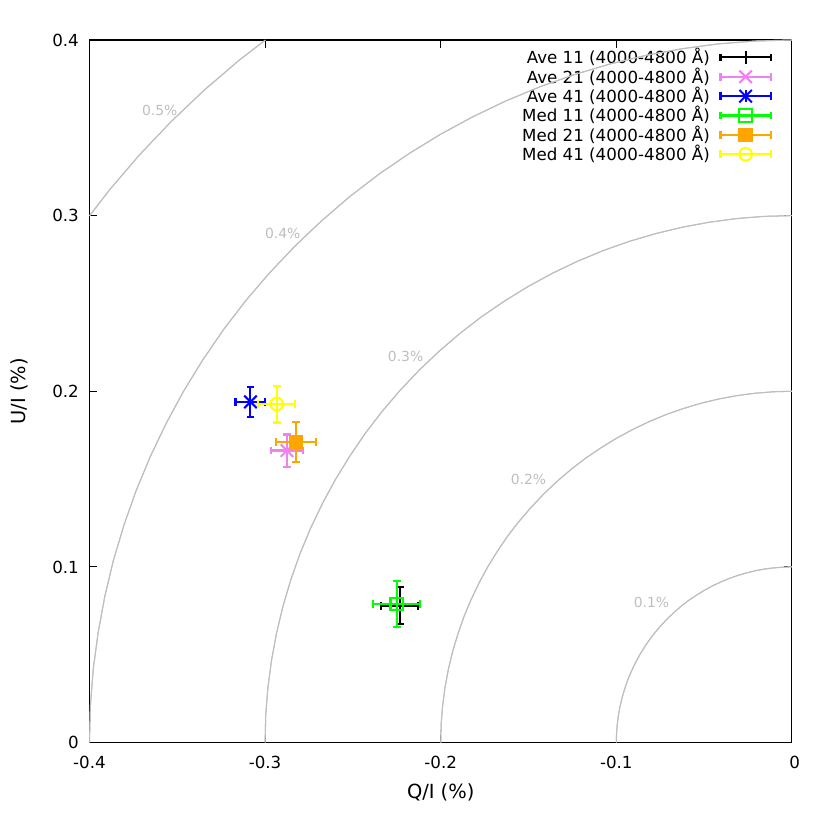}
    \includegraphics[width=\linewidth,trim={0cm 0.5cm 0cm 0cm}]{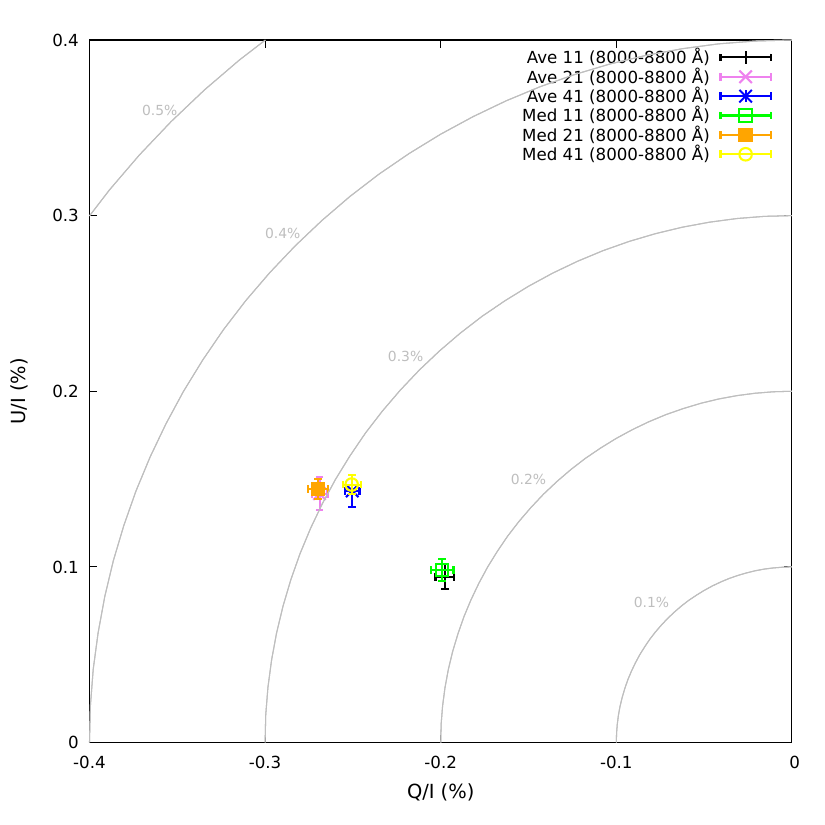}    
    \caption{NGC~1566's Stokes parameter spectra (Q/I and U/I) extracted using different slit lengths and combination methods (average and median), integrated over continuum bands free of strong emission lines. Top panel: 300V grism data. Bottom panel: 300I grism (+ OG590 filter).}%
    \label{Fig:AvgMed_Q_U}%
\end{figure}

We now investigate the differences produced by the two methods on the $q$ and $u$ Stokes parameters. In contrast to Fig.~\ref{Fig:AvgMed_Flux}, there is not enough S/N per resolution element in polarimetry to individually compare the nine $q$ and $u$ spectra, so we computed waveband-integrated $q$ and $u$ values and plotted them, as a function of the method, slit length, and grism in Fig.~\ref{Fig:AvgMed_Q_U}. We see that, within the error bars, all $q$,$u$ combinations are consistent between the methods, independently of the grism and slit length. We note that $P$ increases with slit length but $\theta$ stays constant within the error bars. This is discussed in Sect.~\ref{Analysis:old_data}.

Taken together, the total flux comparison (Fig.~\ref{Fig:AvgMed_Flux}) and the waveband-integrated polarization analysis (Fig.~\ref{Fig:AvgMed_Q_U}) show that median and averaged combinations produce mutually consistent results. Since both methods are statistically equivalent in practice, but the median estimator is less sensitive to outliers -- which can be significant, given that the observed polarization is very low ($<$ 0.4\%) -- we adopted the median combination throughout the rest of the analysis. We did, however, verify that our derived results do not change significantly with the averaging method.

\subsection{Interstellar polarization}
\label{Observation:ISP}

\begin{figure}
    \centering%
    \includegraphics[width=\linewidth]{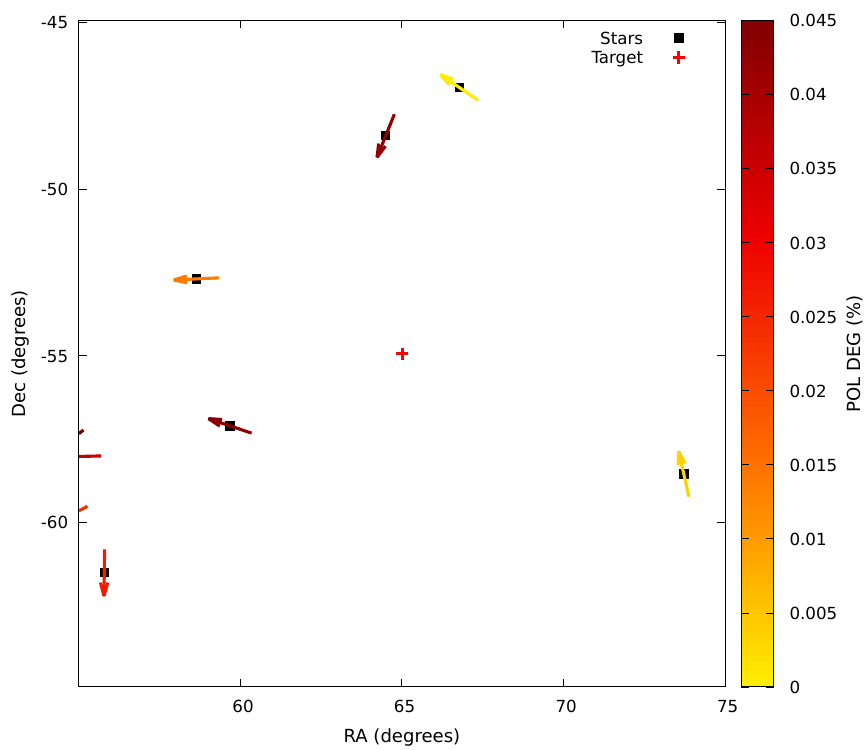}
    \caption{Interstellar polarization (ISP) pattern around NGC~1566, based on star polarization from the compilation of \citet{Panopoulou2025}. The red cross at the center marks the position of the AGN, while the black squares mark the position of various stars within a box with sides of 20 degrees centered around the AGN. Arrows indicate the stellar polarization vectors, with color encoding their polarization degree and orientation corresponding to their polarization angle.}%
    \label{Fig:ISP}%
\end{figure}

The Milky Way's interstellar dust grains aligned by magnetic fields or radiation pressure induce the polarization of light passing through it by differential extinction, a phenomenon called interstellar polarization (ISP). Because non-blazing AGNs are often weakly polarized in the optical band \citep{Antonucci1993}, caution must be taken that the source signal is not too strongly affected by ISP. To verify this, we used the compilation of stars polarization from \citet{Panopoulou2025} and plotted the position in the sky of NGC~1566 in Fig.~\ref{Fig:ISP}, in addition to the polarized starlight detected within a box with sides of 20 degrees centered around the object. As we can see, the ISP nearby NGC~1566 is extremely low ($\le$ 0.05\%) and apparently randomly oriented. The closest star from the AGN is GaiaDR2~4682863148465721728, at a distance of 3.68 degrees from NGC~1566, and it shows a polarization of 0.044\% $\pm$ 0.007\%, associated with a polarization angle of 71.9$^\circ$ $\pm$ 4.5$^\circ$. This value is also confirmed by the maps of ISP of starlight presented by \citet{Mathewson1970}. Both show that the polarization of the stars measured closest to our target are apparently random and negligible, so that ISP is unlikely to strongly affect our polarized results.

\section{A spectropolarimetric analysis}
\label{Analysis}

For the remainder of this paper, we only show spectra extracted with a 1"-wide, 2.75"-long (11 pixels) slit, in order to focus on the AGN component and avoid too much contamination from the host galaxy. The results presented hereafter remain similar for longer extraction apertures.

\subsection{Observed data}
\label{Analysis:Data}%

\begin{figure*}
    \centering%
    \includegraphics[width=\linewidth]{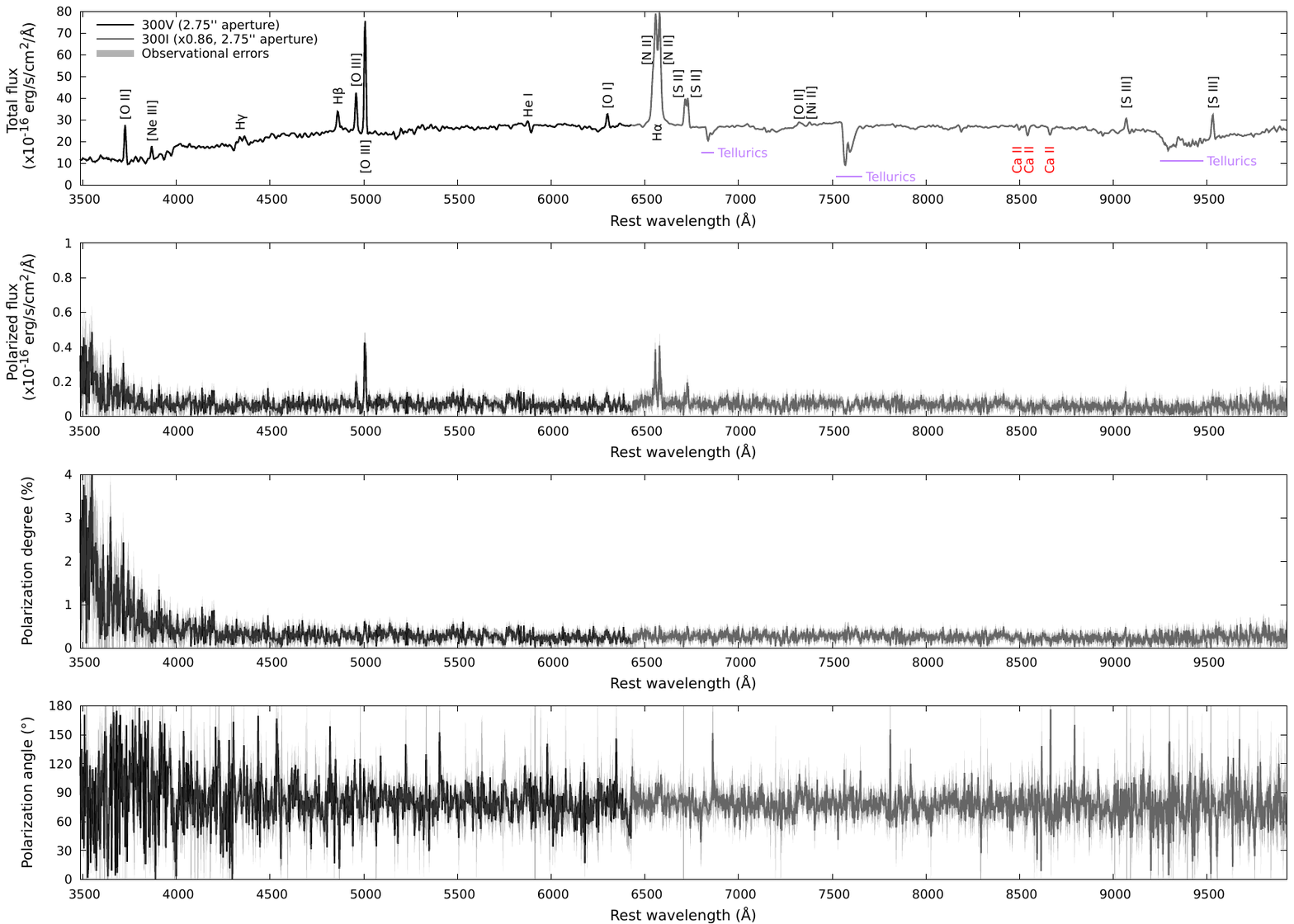}%
    \caption{VLT/FORS2 spectropolarimetry of NGC~1566. Top panel: Total flux spectrum (in 10$^{-16}$ erg~s$^{-1}$~cm$^{-2}$~\AA$^{-1}$). The most prominent telluric absorption lines are indicated in purple, the calcium triplet absorption lines resulting from host starlight are indicated in red, and the most prominent emission lines are labeled in black. The 300I spectrum has been offset to match the 300V spectrum by applying a correction factor of 0.86 to account for the different observing conditions and grisms. Second panel: Polarized flux; that is, the multiplication of the total flux with the polarization degree, $P$. Third panel: Linear polarization degree, $P$. Bottom panel: Polarization position angle, $\theta$. Spectra are shown at native spectral resolution (no binning). Observational errors are indicated in transparent gray for each spectral bin.}%
    \label{Fig:Data}%
\end{figure*}

We present in Fig.~\ref{Fig:Data} the results of the averaged nine spectropolarimetric exposures of NGC~1566 using the 300V and 300I grisms. Focusing on the total flux first (top panel), we see that the spectrum extends from about 3487~\AA\, to 9924~\AA\, in the rest frame. The 300V (in black) and 300I (in dark gray) spectra are slightly offset with respect to each other, which can be due to different observing circumstances (seeing, presence of cirrus...), a logical consequence of observations spanning over two months. We thus applied a multiplicative factor of 0.86 to the 300I spectrum so that the overlapping region between the two grisms (6160-6470~\AA) matches. As labeled on the total flux spectrum panel, we identified the main telluric absorption lines using violet horizontal bars and the calcium triplet -- a key signature of host starlight contamination -- in red. Those features are unrelated to the AGN itself. In black, we identified the main emission lines. The total flux spectrum of NGC~1566 presents no peculiar characteristics with respect to regular Seyfert-2 spectra \citep{Francis1991}: a slightly reddened continuum, indicating modest local dust extinction -- as confirmed by the continuum-subtracted, line-fit H$\beta$/H$\alpha$ ratio of 4.4, entirely compatible with the 4.2 $\le$ H$\beta$/H$\alpha$ $\le$ 4.7 measurements from the literature on NGC~1566 \citep{Martin1974, Osmer1974,Hawley1980} -- and bright narrow emission lines. The permitted H$\alpha$ and H$\beta$ lines appear narrow upon a simple visual inspection. Fitting them using Lorentzian profiles, after subtracting a power-law continuum component and polluting narrow lines (such as [N~II]), gives a FWHM of 777~km~s$^{-1}$ $\pm$ 91~km~s$^{-1}$ for H$\alpha$ and a FWHM of 925~km~s$^{-1}$ $\pm$ 123~km~s$^{-1}$ for H$\beta$. Both show a minor blueshift, with $v_{\rm H\alpha}$ = -266~km~s$^{-1}$ $\pm$ 46~km~s$^{-1}$ and $v_{\rm H\beta}$ = -21~km~s$^{-1}$ $\pm$ 41~km~s$^{-1}$. All the forbidden lines are narrow. The total flux integrated between 4000~\AA\, and 4800~\AA\, amounts to (20.36 $\pm$ 0.31) $\times$ 10$^{-16}$ erg~s$^{-1}$~cm$^{-2}$~\AA$^{-1}$ (accounting for the 300I grism spectral adjustment), while the total flux between 8000~\AA\, and 8800~\AA\, equates to (26.59 $\pm$ 0.40) $\times$ 10$^{-16}$ erg~s$^{-1}$~cm$^{-2}$~\AA$^{-1}$. Compared to the values showed by \citet{Oknyansky2020}, NGC~1566 appears almost ten times fainter than in 2018-2019, consistent with its current type-2 spectral classification.

The polarized flux of NGC~1566 -- that is, the multiplication of the total flux spectrum with $P$ -- is shown in Fig.~\ref{Fig:Data} (second panel). The continuum appears flatter than in total flux, except for a sharp rise at the blue end of the spectrum, reminiscent of the 2018 spectrum of the source \citep{Oknyansky2019}, which was then classified as a type-1 AGN. Bright narrow, forbidden emission lines appear in polarized flux: the [O~III]~$\lambda\lambda$5007,4959 doublet, the [N~II]~$\lambda\lambda$6548,6583  doublet, the [S~II]~$\lambda\lambda$6717,6731 doublet and [S~III]~$\lambda$9068. There are no H$\alpha$ and H$\beta$ lines in polarized flux, at least within the uncertainties, which supports a disappearance of the region rather than obscuration, as scattering should preserve the broad line signatures in polarized flux even when the BLR is hidden in total flux. The presence of narrow emission lines in the polarized flux indicate that they are polarized too. If intrinsically unpolarized, these lines would depolarize the continuum and leave the polarized flux essentially flat at their wavelengths. Their clear detection in polarized flux instead suggests that the narrow line emission is polarized in a manner similar to the continuum, likely through the same scattering mechanism. Physically, this means that the scattering region is extended, so that it can intercept both nuclear continuum photons and narrow line region (NLR) photons. 

The polarization degree (Fig.~\ref{Fig:Data}, third panel) is very informative. The linear continuum shows a polarization that is wavelength-independent over 4000-9900~\AA, confirming the absence of significant ISP and advocating for electron scattering as the dominant polaro-genesis mechanism. However, we note that $P$ strongly evolves shortward of 4000~\AA. The continuum sharply rises to $\sim$ 2\%, while it was 0.24\% $\pm$ 0.01\% between 4000~\AA\, and 4800~\AA\, and 0.22\% $\pm$ 0.01\% between 8000~\AA\, and 8800~\AA. It could be associated with the strong rise of the blue continuum observed in 2018 \citet{Oknyansky2020}, observed as an echo in polarized flux. However, this rise in $P$ should be regarded as purely qualitative, as the statistical significance degrades markedly in this spectral region, which lies near the blue edge of the 300V grism where the sensitivity drops. In the optical/near-IR part of the spectrum, the polarization signal measured in the aforementioned bright emission lines is consistent with that of the continuum, indicating that the narrow line photons share the same polarization state. This suggests that the forbidden line emission is polarized through scattering, rather than being intrinsically polarized at the site of emission, as expected for UV-photoionized gas in the NLR.

Finally, the polarization position angle is shown in the bottom panel of Fig.~\ref{Fig:Data}. $\theta$ appear wavelength-independent over the 4000-9900~\AA\, waveband, confirming that only one polarization process dominates the production of polarized light. The polarization angle becomes less well defined shortward of 4000~\AA\, for the same reasons explained before, but stays consistent with the averaged value. Integrated over the 4000–4800~\AA\, waveband, $\theta$ = 80.4$^\circ$ $\pm$ 1.3$^\circ$, which is very similar to the 8000–8800~\AA\, value (77.3$^\circ$ $\pm$ 0.9$^\circ$). As is described in Sect.~\ref{Introduction}, there are no radio position angles to compare the optical polarization angle to, in order to infer the location and geometry of the dominant scattering mechanism. If we use the [O~III] and X-ray position angle as a proxy for the polar axis of the AGN \citep{Elvis1990,daSilva2017}, i.e. 120$^\circ$-130$^\circ$, we find that the optical polarization angle of NGC~1566 is neither parallel nor perpendicular to this axis. This supports there being an asymmetric distribution of scatterers in the circumnuclear region.

\subsection{Comparison to previous polarimetric observations}
\label{Analysis:old_data}%

As was already mentioned in Sect.~\ref{Introduction}, NGC~1566 polarization was only measured twice in the past. The April 20, 1980, 3800–5600~\AA\, band observation reported by \citet{Martin1983} while the source was in a type-2 state is 0.55\% $\pm$ 0.24\% in a 4" aperture (uncorrected for starlight and interstellar contamination). The associated polarization angle, $\theta$, is 52.6$^\circ$ $\pm$ 11.6$^\circ$. Interestingly, compared to the putative AGN polar axis traced by the NLR and extended X-ray structure, \citet{Martin1983}'s polarization angle is nearly perpendicular, as is the case for all Seyfert-2s \citep{Antonucci1993}. While $P$ is compatible with our measurement within the errors bars, the value of $\theta$ is different at the 2.4$\sigma$ level. The significance is moderate and not decisive here.

The second polarization campaign, and the last one known to us, comes from \citet{Felton1999}'s thesis, in which V-band measurements obtained on February 5, 1997 and R-band measurements acquired on February 15, 1997 at the Cassegrain focus of the South African Astronomical Observatory (SAAO) 1-meter telescope are reported. The disclosed V-band (central wavelength: 5450~\AA\,, FWHM of 130~\AA) nuclear aperture polarization of NGC~1566 was 1.33\% $\pm$ 0.18\% at 46$^\circ$ $\pm$ 4$^\circ$ in a 4" aperture. In the R band (central wavelength: 6613~\AA\,, FWHM of 70~\AA), and for the same aperture, the polarization was 3.25\% $\pm$ 0.27\% at 50$^\circ$ $\pm$ 2$^\circ$. Here again, with better precision than for \citet{Martin1983}'s measurement, the extended, polar X-ray structure reported by \citet{Elvis1990} and the optical/NIR polarization angle are perpendicular. This is particularly unusual, because, at that time, NGC~1566 was in the type-1 state and the unified model of AGN stipulates that the majority of Seyfert-1s have their polarization angle parallel to that of their radio/NLR structure \citep{Antonucci1993}. In all cases, neither the polarization degree nor the polarization angle reported by Felton are compatible with our measurements at more than 6$\sigma$. It is thus safe to conclude that both $P$ and $\theta$ have significantly varied between 1997 and 2025.

The manuscript of \citet{Felton1999} also reports polarimetric measurement in the V and R band for smaller and larger apertures. She found that $P$ decreases with increasing aperture (from 1.47\% at 3" to 1.04~\% at 6" in the V band, and from 3.63\% at 3" to 2.61~\% at 6" in the R band), while $\theta$ stays constant within the error bars (see Tab.5.13 in \citealt{Felton1999}). This is logical as, with increasing apertures, more unpolarized starlight dilutes the observed polarization. This is in contradiction with what we found in Sect.~\ref{Observation:AvgMed}; see Fig.~\ref{Fig:AvgMed_Q_U}. In our VLT/FORS2 observation, $P$ increases with a larger slit length, while $\theta$ stays constant. Compared to the data of \citet{Felton1999}, this requires additional polarized flux on arcsecond scales that was not present (or much weaker) in 1997. It strongly suggests a new or enhanced extended scattering component -- the echo from the 2018 activity scattering onto the LNR, or a change in the geometry and/or illumination of circumnuclear structures -- coherent with the disappearance of the broad emission lines, flux fading, and low polarization of NGC~1566.

The fact that NGC~1566 becomes more polarized when brighter (in a type-1 state, with broad emission lines) is counterintuitive. According to the standard model of AGNs \citep{Antonucci1993}, in a radio-quiet, pole-on object, we have an unobstructed view of the direct light from the AGN continuum, plus the direct BLR emission and only a small contribution from light that have scattered on the inner parsec components (accretion flow, BLR, torus inner wall...). It is known that the direct continuum is essentially unpolarized \citep{Gaskell2012}. Scattered light, on the other hand, is polarized parallel to the radio structure axis, but usually contributes only a tiny fraction of the total flux \citep{Marin2012}. Therefore, under standard AGN polarimetry expectations, when the nucleus brightens and the direct (unpolarized) light dominates, the observed polarization should decrease, due to dilution; this is why Seyfert-1s typically have lower polarization (0.1-2\%) than Seyfert-2s (1–10\%), see \citet{Marin2014}. But, in NGC~1566, the opposite happens and it is counterintuitive. It implies that, when the source transitions from a type-2 to a type-1 classification, one of the following must occur: 1) either the polarized (scattered) component increases faster than the direct continuum; 2) a new scattering region becomes visible or more efficiently illuminated; or 3) the direct line of sight to the nucleus is partially obstructed. In the latter case, the relative contribution of scattered light is enhanced, implying that the unobstructed flux should be even higher than observed. The last scenario appears the least plausible of the three, since the broad lines should still be detected in polarized light and since the extinction toward NGC~1566 is only moderate (see Sect.~\ref{Analysis:Data}). 

Another intriguing point is that the polarization reported by \citet{Felton1999} is strongly wavelength-dependent, while ours is not. The author declared that, “this rise in polarization level toward longer wavelengths may be because the diluting unpolarized flux from the host galaxy is less at these wavelengths, so that the polarized nuclear emission constitutes a larger fraction of the observed flux.” This is what we now investigate.

\subsection{Host galaxy dilution}
\label{Analysis:Host}%

In order to measure the true, intrinsic AGN polarization, it is necessary to remove the diluting contribution of the host galaxy. To do so, we used the \textsc{starlight} code \citep{CidFernandes2005}, a nonparametric spectral synthesis code that fits an observed spectrum in terms of a combination of stellar population spectra, allowing also for dust reddening and kinematics. 
The stellar population models come from an updated version of the \citet{Bruzual2003} models (see \citealt{Martinez2023}). The specific base used here span ages from 0 to 14 Gyr and metallicities in the 0.2 to 3.5 solar ranges. Reddening ($A_V$ = 0.6 mag from the fits) was modeled with a \citet{Calzetti2000}'s law. We complemented this base with a set of $F_\lambda \propto \lambda^\beta$ power laws to account for the AGN continuum. Telluric features were masked from the fit. Because the instrumental resolution differs slightly between the blue and red spectral ranges, we homogenized the effective velocity dispersion across the full wavelength range prior to the fit and adopted a second, refined emission-line mask to better exclude weak features.

\begin{figure}
    \centering%
    \includegraphics[width=\linewidth,trim={0cm 0.5cm 0cm 0cm}]{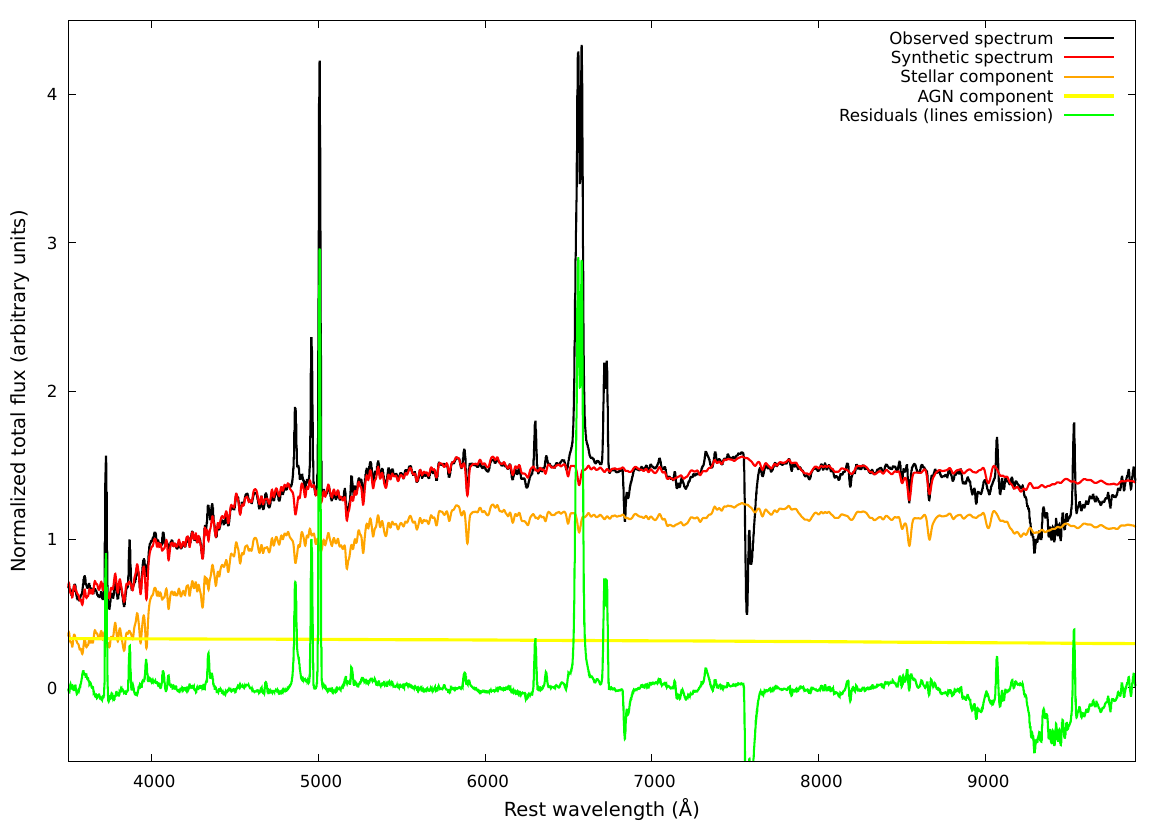}%
    \caption{Fits to the observed spectrum of NGC~1566 using the \textsc{starlight} code \citep{CidFernandes2005}. In black is the observed total flux spectrum (as shown in Fig.~\ref{Fig:Data}, top panel), in red is the synthetic spectrum obtained with the code (that is composed of the stellar component in orange and of the AGN component in yellow), and in green are the residuals. Negative fluxes in the residuals are due to the uncorrected telluric absorption lines.}
    \label{Fig:Host}%
\end{figure}

The results of the fit are presented in Fig.~\ref{Fig:Host}. They indicate that the observed total flux continuum is predominantly constituted of starlight emission (shown in orange), with a wavelength-dependent contribution that varies from about 50\% at the near-UV end to 78\% in the near-IR. The almost flat AGN continuum (shown in yellow) is the result of a reddened $\lambda^{-0.5}$ power-law. We verified that using an off-nuclear spectrum as an alternative stellar template (see, e.g., \citealt{Shapovalova2019}) yields a comparable fit quality and does not significantly modify the inferred AGN continuum fraction. Once the stellar and AGN continua are subtracted from the observed spectrum, only the emission lines are left (in green in Fig.~\ref{Fig:Host}). Here again, no broad components are detected in the permitted lines.

\begin{figure*}
    \centering%
    \includegraphics[width=\linewidth,trim={0cm 0.2cm 0cm 0cm}]{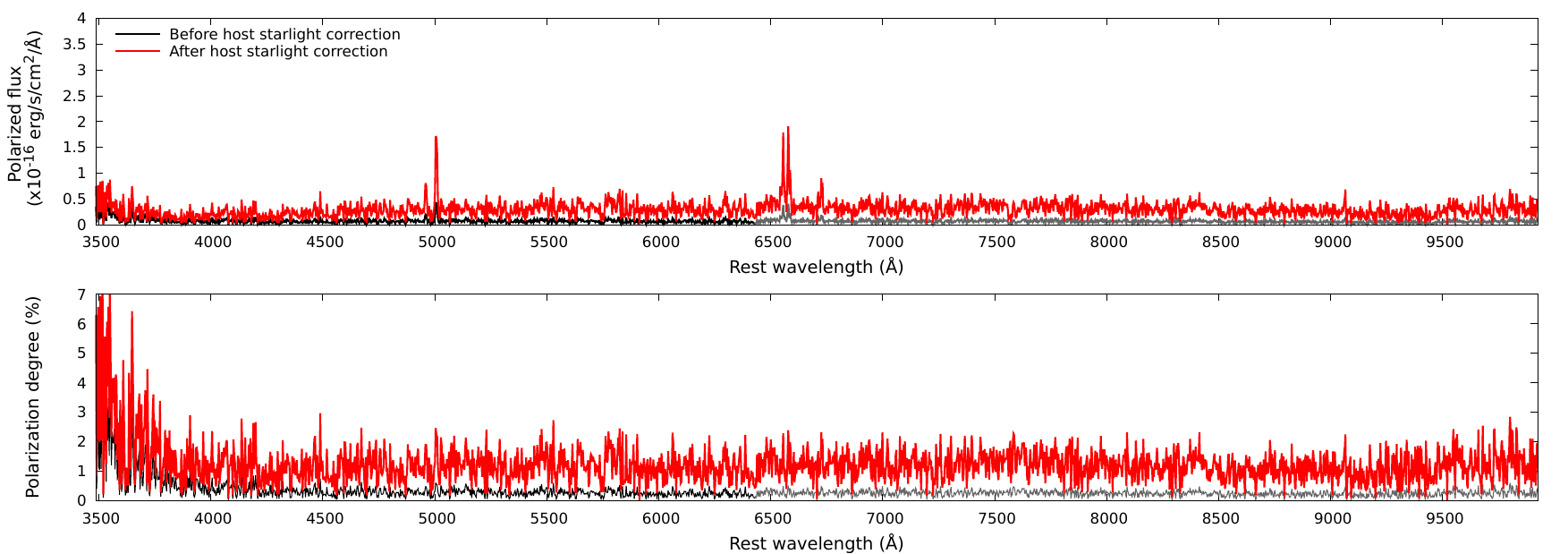}%
    \caption{Polarized flux (top panel) and polarization degree (bottom panel), prior and after the removal of the unpolarized starlight contribution evaluated with the \textsc{starlight} code, in black and red, respectively.}%
    \label{Fig:Data_corrected}%
\end{figure*}

If we suppose that starlight is essentially unpolarized \citep{Kemp1987}, we can remove the Stokes I component of the host emission to our polarized spectrum, in order to recompute the $q$ and $u$ components, and thus the corrected polarization degree. Under this hypothesis, the polarization angle remains unchanged. We present the results of such a subtraction in Fig.~\ref{Fig:Data_corrected}, where we show the diluted -- in black -- and host-corrected -- in red -- polarized flux (top) and polarization degree (bottom). The correction we applied did not change the shape of either spectrum, confirming that electron-scattering is the dominant polarization mechanism producing the observed polarized flux. H$\alpha$ and H$\beta$ lines are still undetected and the rise in polarized flux in the near-UV band appears even stronger. The host-corrected polarization reaches about 1\%, indicating that NGC~1566 is most certainly seen pole-on, as edge-on AGN intrinsic polarization is largely in excess of 1\% \citep{Marin2012}. Importantly, the intrinsic polarization remains weak ($\lesssim$1-2\%) for all reasonable stellar subtraction methods tested, implying that the changing-look behavior of NGC~1566 is likely driven by genuine structural changes in its inner accretion flow, rather than edge-on obscuration from dense, partially ionized (or dusty) gas clouds originating from the BLR (torus).

Such analysis disfavors the argument of Felton. The rise in polarization that she observed between the green and red bands cannot be due to a lower starlight contribution, as the host fractional emission is smaller in the V band than in the R band (as opposed to her suggestion). This explanation would only be viable if the AGN continuum had been significantly redder in 1997, thereby increasing its relative weight in the red band. However, NGC 1566 was in a clear type-1 state at the time -- with broad emission lines and no indications of strong extinction \citep{Felton1999} -- making such a scenario unlikely. As a result, it is plausible that the rise in $P$ with wavelength observed in 1997 was intrinsic to the AGN, which suggests dust scattering processes rather than Thomson scattering as seen in 2025. This seems in agreements with the fact that, in 1997, NGC~1566's polarization was dominated by perpendicular scattering onto the NLR.

\section{Variability and the binary SMBH hypothesis}
\label{Variability}

Because our observing campaign lasted for about two months and that we successfully achieved nine exposures per grism, we are able to investigate the variability in NGC~1566, both in terms of total flux spectra and integrated polarization. However, the individual polarized spectra are unusable, as they individually lack statistics.

\subsection{Total flux spectra}
\label{Variability:Flux}%

\begin{figure*}
    \centering%
    \includegraphics[width=0.5\linewidth,trim={0cm 0.5cm 0cm 0cm}]{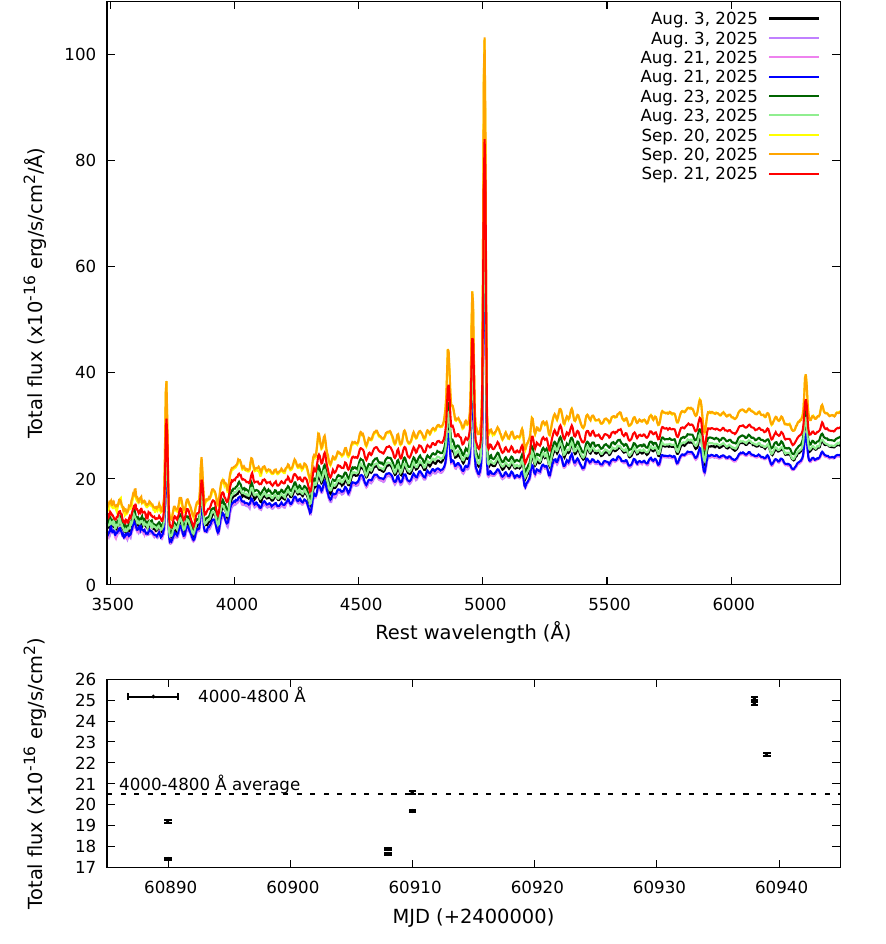}%
    \includegraphics[width=0.5\linewidth,trim={0cm 0.5cm 0cm 0cm}]{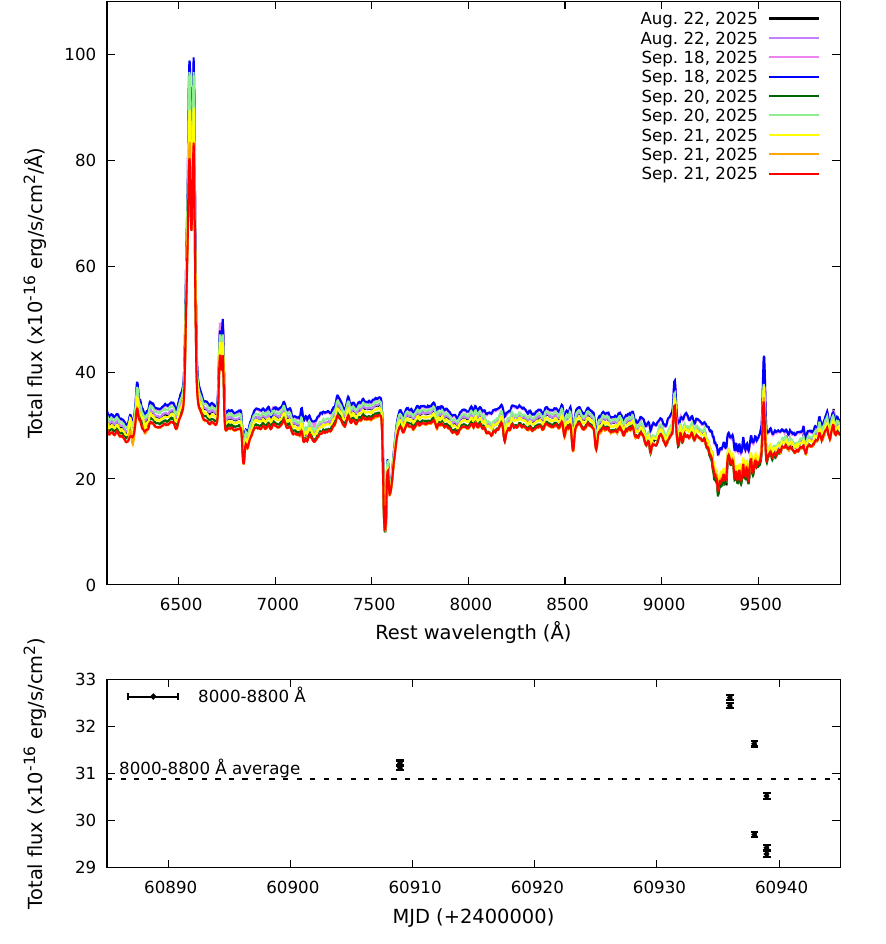}%
    \caption{Variability in total flux of the nine observations of NGC~1566 taken between August and September 2025, for the 300V grism (left) and the 300I grism + OG590 filter (right). Top panels: Spectra in $10^{-16}$~erg~s$^{-1}$~cm$^{-2}$~\AA$^{-1}$, color-coded by observing date. Bottom panels: Evolution of the flux, integrated over different bands, as a function of the modified Julian date (MJD). The average flux is indicated with a dashed line in each bottom panel.}%
    \label{Fig:Var_FLux}%
\end{figure*}

We begin by investigating the spectral shape variations over August-September 2025. We plot in Fig.~\ref{Fig:Var_FLux} the nine 300V (top left) and 300I (top right) spectra, as well as their broadband, 4000-4800~\AA\, (bottom left) and 8000-8800~\AA\, (bottom right) integrated total flux as a function of the observing date. 

We find that the flux levels vary by about 20\% between the faintest and the brightest observations, with up to 7\% variability within hours, which is perfectly consistent with normal Seyfert-1 behaviors \citep{Peterson2001}, but inconsistent with type-2 objects, which usually show slow (months to years) and low-amplitude (a few percent at best) variability in the optical band \citep{Yip2009}. It is another indication that NGC~1566 is viewed close to its polar axis, rather than along the equatorial plane. Plotting the difference spectra (spectrum at a given MJD minus the averaged spectrum) and the ratio between each exposure combination, none show statistically significant changes in the emission line profiles; they rather vary in intensity. Only a very slight variation in the slope of the 300V continuum -- which becomes bluer -- is observed. 

\begin{figure*}
    \centering%
    \includegraphics[width=\linewidth]{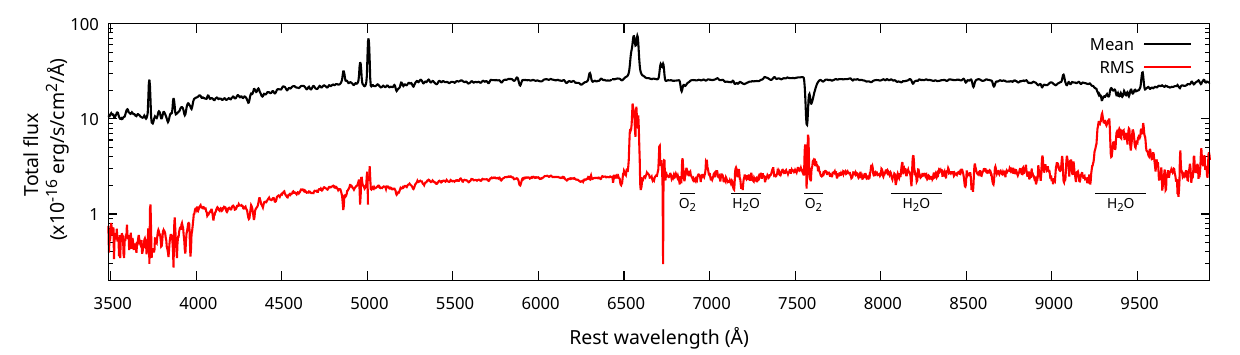}%
    \caption{Mean (in black) and rms (in red) spectra of our nine observations.}%
    \label{Fig:RMS}%
\end{figure*}

Figure~\ref{Fig:RMS} shows the root-mean-square (rms) spectrum of the observations after normalizing the spectra using narrow forbidden lines ([O III]~$\lambda$5007 in the 300V grism and the [S II]~$\lambda\lambda$6716,6730 doublet in the 300I grism). The rms spectra are used to highlight where variability is the strongest. They usually show prominent dips when emission components vary less than the continuum, and stronger variability in emission components produces emission peaks (see, e.g., \citealt{Vincentelli2025}). We observe that our rms spectrum shows a continuum shape almost identical to the mean spectrum, confirming that the main variability is due to scaling factors such as seeing, variable atmospheric extinction, or small differences in the object centering into the slit. Residual peaks remain at the positions of narrow emission lines, likely due to slight differences in slit losses or wavelength sampling. At the locations of the broad Balmer lines H$\beta$, H$\gamma$, and H$\delta$, small dips are observed, consistent with the absence of intrinsic variability in the aformentionned lines during this observing campaign. In the H$\alpha$ region, the rms is dominated by the [NII] blend, preventing a clear detection of (putative) variability. In the red band, telluric absorption, which is intrinsically variable, is observed in different bands indicated in the graph and the narrow dip observed at the [S II]~$\lambda\lambda$6716,6730 doublet doublet arises as an artifact of our rms computation, caused by the limited spectral resolution of the observations. Overall, the rms spectrum does not reveal any significant traces of even a very weak BLR component and indicates that variability between August and September 2025 is mainly driven by varying observing conditions rather than intrinsic changes in the BLR.

\subsection{Polarization}
\label{Variability:P}%

\begin{figure*}
    \centering%
    \includegraphics[width=0.48\linewidth]{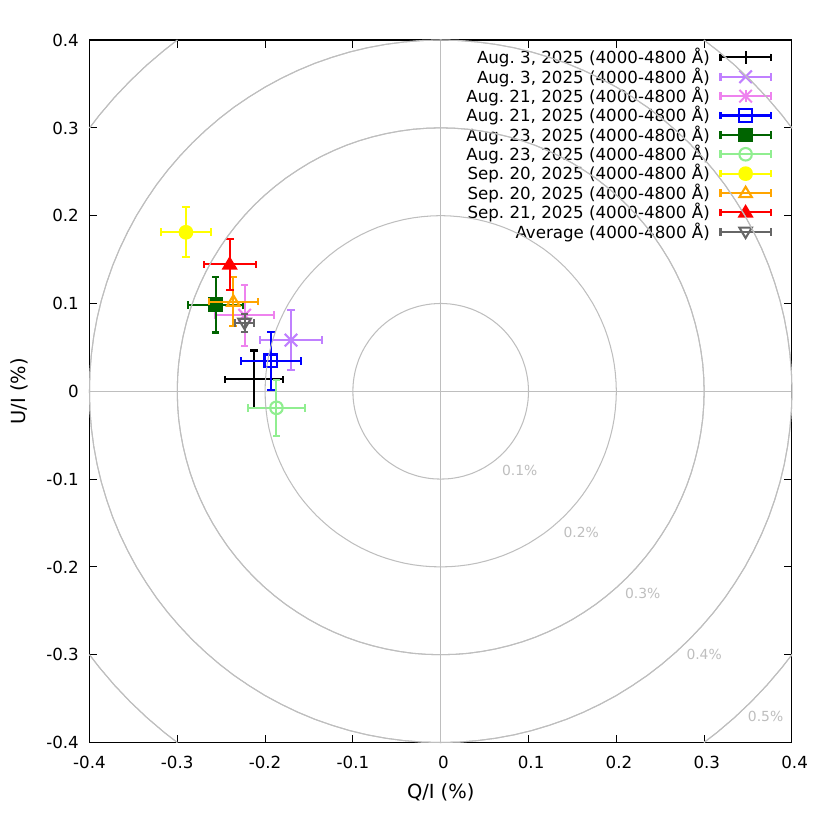}    
    \includegraphics[width=0.48\linewidth]{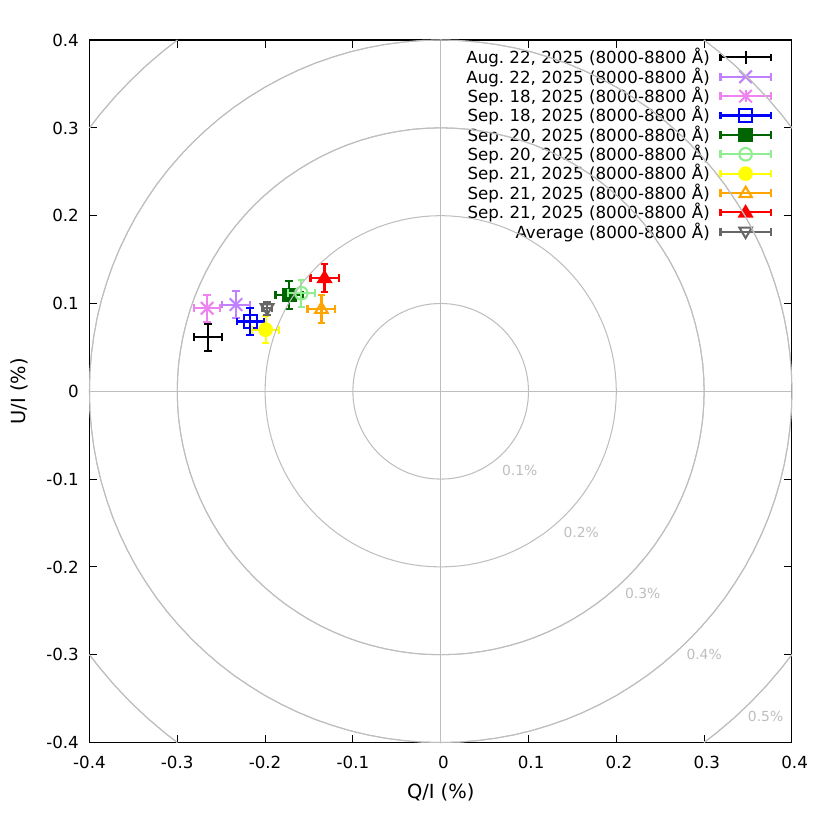}
    \caption{Variability in polarization for the 300V (left) and 300I (right) grisms for the nine observations of NGC~1566 taken between August and September 2025. The panels show the Q/I versus U/I diagrams, with each observation color-coded.}%
    \label{Fig:Var_P_V}%
\end{figure*}

\begin{figure}
    \centering%
    \includegraphics[width=\linewidth]{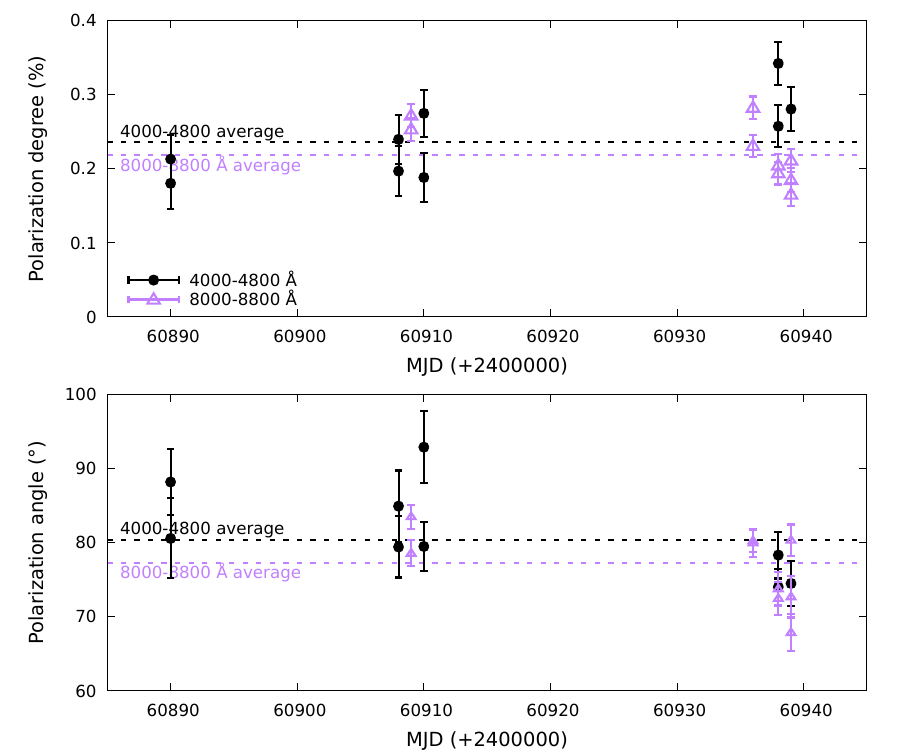}%
    \caption{Evolution of the polarization degree $P$ (top panel) and polarization angle (bottom panel), integrated over 4000--4800~\AA\, in black, and over 8000--8800~\AA\, in purple, as a function of the MJD. The dashed lines indicate the average polarizations, computed from all nine merged observations (as in Fig.~\ref{Fig:Data}).}%
    \label{Fig:Var_P_avg}%
\end{figure}

Our second test concerns the wavelength-integrated polarization of NGC~1566 in the 4000--4800~\AA\ and 8000--8800~\AA\, wavebands. We present in Figs.~\ref{Fig:Var_P_V} and \ref{Fig:Var_P_avg} the variability in polarization for the nine observations of NGC~1566 taken between August and September 2025. Figure~\ref{Fig:Var_P_V} uses Q/I versus U/I diagrams, since polarization is a vector quantity, best represented in the Stokes plane where both its amplitude and position angle can be simultaneously visualized.

Figure~\ref{Fig:Var_P_V} (left), which focuses on the 300V grism, shows that the measurements are aligned along a straight line with a very small dispersion around it, and a clear shift from the origin of the QU diagram. The averaged point (taken from Sect.~\ref{Analysis}) is located in the middle of the cloud of data points. The polarization degree is weak for all exposures, but varies by more than 0.15\% in both Q/I and U/I (to be compared with the instrumental polarization, that is inferior to 0.1\%). We observe the same behavior in the 300I grism (Fig.~\ref{Fig:Var_P_V}, right), but with points aligned along an axis that, even if it does not pass through the origin of the diagram too, appears perpendicular to the line of points observed at 300V. The polarization also varies in Q/I but appears almost constant in U/I. The averaged point is also situated in the center of the cluster of the nine measurements. The 300V data points are, on average, higher in polarization than the 300I (see Fig.~\ref{Fig:Var_P_avg}), which can be directly linked to the lesser impact of starlight dilution from the host (see Fig.~\ref{Fig:Host}).

The 300V and 300I data points, being aligned along a line that is straight but that does not pass through the origin of the QU plane, indicate that a single variable component dominates the polarization, added to a constant offset. This constant polarization, being relatively small, could very well correspond to the sum of the interstellar, host, and instrumental polarizations. Regarding the variable component, because the $q$,$u$ points have a relatively small dispersion around this line, the variable component is clearly astrophysical in origin and not caused by noise, instrumental effects, or scintillation originating from atmospheric turbulence. What is particularly noteworthy is that the 300I points are also aligned on a straight line -- which also does not pass through the origin of the QU plane -- but this line is roughly perpendicular to the direction of the 300V points. This implies that the direction of variation in QU is different between the two bands, suggesting scattering processes in two distinct regions of the AGN, perpendicular to each other. It may indicate that part of the AGN polarization we observe comes from scattering inside the NLR (echoing the past 2018 outburst, as revealed in the blue polarized flux), while a larger fraction of the flux comes from the central engine (accretion flow or a BLR residue). The rapid, small-amplitude variations that we observe in the QU plane (about 0.05–0.2\%, sometimes within a day) are more consistent with an asymmetric distribution of scatterers, rather than with binary SMBH activity, unless the system is exceptionally tight.

\section{Discussion}
\label{Discussion}

Spectropolarimetry has revealed a complex and evolving geometry in NGC~1566. In its current, 2025 state, it exhibits at least two distinct and perpendicular scattering regions: one associated with the core engine along the equatorial plane, and another one along the polar axis, probably linked to the NLR, with significant changes over timescales of decades. The counterintuitive behavior of the polarization, as revealed in Sect.~\ref{Analysis:old_data}, is key here. The increase in polarization during bright (type-1) phases cannot be explained by the standard model \citep{Antonucci1993} and requires either a variable scattered component that grows faster than the direct continuum (reminiscent of the yet-to-be-solved Alloin problem, whereby emission lines react before the optical continuum to variations in the ionizing flux, \citealt{Alloin1985}), or the emergence of a new scattering region during bursts. Interestingly, \citet{Oknyansky2024} also theorized that strong luminosity outbursts may substantially modify the dust distribution in the nuclear environment, leading to temporary dust sublimation followed by a gradual reformation during subsequent low states. If such dust is not confined to the classical torus but is also present in polar regions (see, e.g., \citealt{Honig2013}), its time-dependent recovery could intermittently obscure the BLR and contribute to the observed changing-look behavior.

In the most commonly accepted changing-look picture, the BLR is assumed to persist as a population of clouds within a few light-days to light-weeks from the black hole, with the visibility of the broad lines primarily governed by changes in the ionizing continuum from the accretion flow (see, e.g., \citealt{Noda2018}). In this framework, decreases in the accretion rate can strongly reduce the photoionization of existing BLR gas without requiring wholesale inflow of new material \citep{Netzer2015}. In our observations, the absence of broad-line signatures in polarized light, combined with the current low and wavelength-independent polarization, runs contrary to simple obscuration of an otherwise normal BLR and instead suggests that the region is either intrinsically very weak or temporarily absent during the present low state.

An eccentric accretion disk hypothesis remains viable, as the observed rapid variations and asymmetric geometry are consistent with a perturbed disk or a turbulent BLR dominated by the disk \citep{Ochmann2024}. Another interesting avenue to explore to explain the recurrent activity is the occurrence of stellar tidal interactions with the central black hole. While a full tidal disruption event would produce a single, dramatic flare, partial disruptions or repeated mass stripping episodes can arise if a star follows a bound, highly eccentric orbit around the black hole. In such a scenario, successive close passages could lead to recurrent bursts of enhanced accretion and luminosity, potentially accounting for the quasi-periodic bright phases observed in NGC~1566 \citep{Campana2015,Komossa2015,Grupe2015}. The distinction between these scenarios unfortunately remains difficult given the lack of polarimetric monitoring of NGC~1566 and the absence of spectropolarimetry obtained during a confirmed type-1 state.

A binary SMBH hypothesis is less favored. As is noted by \citet{Kollatschny2024}: “a nearly edge-on binary orientation is necessary to produce self-lensing systems of massive black hole binaries,” and our analysis clearly indicates that NGC~1566's core is seen close to pole-on. While a self-lensing binary does not strictly require alignment between the binary orbital plane and the NLR axis, such a configuration would imply a strong misalignment between the orbital plane and the AGN symmetry axis, which appears unlikely given the polar viewing angle inferred for NGC~1566. In addition, the observed polarization variations (short timescales and small amplitudes) are better explained by an asymmetric distribution of scatterers \citep{Marin2017} than by a binary system, unless the latter is exceptionally compact. Indeed, \citet{Savic2019} have shown that binary SMBHs produce unique variations in $\theta$, with periodic rotations of several tens of degrees around Balmer emission line peaks, depending on the separation between the two black holes and the viewing angle. Such variations are not seen in our spectra but, for completeness, we present in Appendix~\ref{Appendix:Savic} the model configuration from \citet{Savic2019} that best reproduces our observation for comparison. Moreover, \citet{Dotti2022} computed the expected polarized signature from one of the binary components by a surrounding circumbinary ring and showed that the variations are extremely small, about 20~\% of the scattered light fraction, leading to polarization degree changes at the $\le$ 0.1~\% level and polarization angle oscillations of only $\sim$ 1$^\circ$. Those values are incompatible with the variations observed in $P$ and $\theta$ revealed by our QU diagrams (Fig.~\ref{Fig:Var_P_V}).

\section{Conclusion}
\label{Conclusion}

We have obtained and analyzed nine spectropolarimetric exposures of NGC~1566 using the VLT/FORS2 between August and September 2025. Our observations indicate that NGC~1566 is currently in a type-2 state, with no broad components detected in H$\alpha$ and H$\beta$ in total flux. This flux is approximately ten times weaker than in 2018-2019. The polarized continuum is surprisingly small (0.22-0.24\% over 4000-9900~\AA) and wavelength-independent, indicating electron scattering as the dominant mechanism. The polarization angle is also constant, at $\theta \sim$ 77-80$^\circ$ across the entire spectral band. We note an apparent strong increase in polarization in the UV, up to 2\% below 4000~\AA. Additionally, there is no trace of a BLR signature visible in polarized flux, suggesting that it has truly disappeared rather than been obscured.

Compared with past observations (1980, 1997), we find a counterintuitive behavior: NGC~1566 is more polarized when it is brighter (type-1 phase), contrary to the predictions of the standard unified model. In addition, archival data show significant changes in $P$ and $\theta$ between 1997 and 2025 at more than 6$\sigma$, with a polarization degree increasing with wavelength in 1997, but flat in 2025. As we extract $P$ at larger spatial scales, our polarized continuum increases, something at odd with usual observations, indicating the presence of an additional polarized flux on arcsecond scales (as recently observed by Ag\'is-Gonz\'alez et al., accepted, in GSN~069). If we fit the spectra using a stellar synthesis model to extract the host galaxy's contribution from our spectra, the contribution of starlight is found to vary from about 50\% in the UV to about 78\% in the near-IR. After correction, the intrinsic linear continuum polarization of the AGN reaches 1\% in the optical/red band, confirming that the polarization signal is dominated by equatorial scattering and thus that the view toward NGC~1566 is near the pole. The 2025's optical polarization angle is neither parallel nor perpendicular to the assumed polar axis ($\sim$ 120-130$^\circ$ according to [O~III] and X-rays mapping), which supports there being an asymmetric distribution of circumnuclear scatterers and multiple scattering regions with different orientations. Uncommonly, the polarization angle of the 1997's observation, as NGC~1566 was a bright, type-1 object, was perpendicular to the polar axis, a behavior more expected of type-2 objects. 

The analysis of the variability of our spectra between August and September 2025 indicates that total flux variations (up to 20\%) between observations most likely originate from observational effects, while the polarization, plotted on a Q/U diagram, shows points aligned on straight lines not passing through the origin and perpendicular between the 300V and 300I grisms. Perpendicular directions of variation between 300V and 300I suggest two distinct scattering regions: one equatorial (probably a scattering disk) and another one situated along the polar axis of the AGN (the NLR). Rapid but small-amplitude variations in polarization (0.05-0.2\%, sometimes within a day) are also observed, more consistent with an asymmetric distribution of scatterers than with a binary system of SMBHs, unless in a very tight configuration.

Overall, our spectropolarimetric campaign of NGC~1566 in a type-2 state strongly disfavors obscuration by dense clouds as the cause of the changing-look nature of this object and is hard to reconcile with a binary SMBH too. However, it supports accretion-driven photoionization and/or structural changes in the internal accretion flow and the BLR. A spectropolarimetric monitoring during a type-2 $\rightarrow$ type-1 transition would be crucial to definitively distinguish between eccentric accretion disks, turbulent disk-dominated BLRs, or repeated partial tidal disruptions, by observing how polarization in the continuum and lines evolve during the reappearance of the BLR, in order to probe its true geometry.

\begin{acknowledgements}
The authors thank the anonymous referee for her/his careful reading of the manuscript and constructive comments, which helped improve the clarity and quality of this work. F.M. and J.B. acknowledge financial support from the french national space agency (CNES) and the CNRS Action Thématique Phénomènes Extrêmes et Multi-messagers (AT-PEM). D.H. is F.R.S-FNRS research director (Belgium). V.O. has been partly supported by Israeli Science Foundation grant no.~2398/19
\end{acknowledgements}

\bibliographystyle{aa}
\bibliography{biblio}

\begin{appendix}
\onecolumn

\section{Comparison to binary SMBH simulations}
\label{Appendix:Savic}

For completeness purposes, we extracted from the publication of \citet{Savic2019} the binary SMBH model that best resemble the shape of the H$\beta$ line\footnote{In the case of H$\alpha$, this is particularly challenging because the line is strongly blended with narrow emission components and host galaxy starlight. Removing these contributions would require assumptions about the intrinsic polarization of the narrow lines. Indeed, narrow forbidden lines may themselves be polarized through processes such as scattering in the narrow-line region or dichroic extinction in intervening dust. Given that the overall polarization level in our data is very low (of order $\sim$ 1\% at most after correcting for host starlight), disentangling these components would introduce uncertainties comparable to or even greater than the signal itself, making such a decomposition unreliable with the present data.} in total flux, as observed in the individual and median spectra of NGC~1566 (see Figs.~\ref{Fig:Data} and \ref{Fig:AvgMed_Flux}). Most of the binary models from the aforementioned publication result in double peaked line profiles with phase-dependent asymmetries in the wings, while single-peaked profiles (as observed in NGC~1566) arise only for a limited range of geometrical configurations (see Figs~A1, A2 and A3 and the associated model in \citealt{Savic2019}). Such model, with two SMBHs of comparable mass at an orbital distance of 47.6 light days, is seen from a type-1 inclination ($\sim$ 18$^\circ$), as expected from NGC~1566.

~\

\begin{figure*}
    \centering
    \includegraphics[width=0.7\textwidth,trim={0cm 0cm 0cm 0cm}]{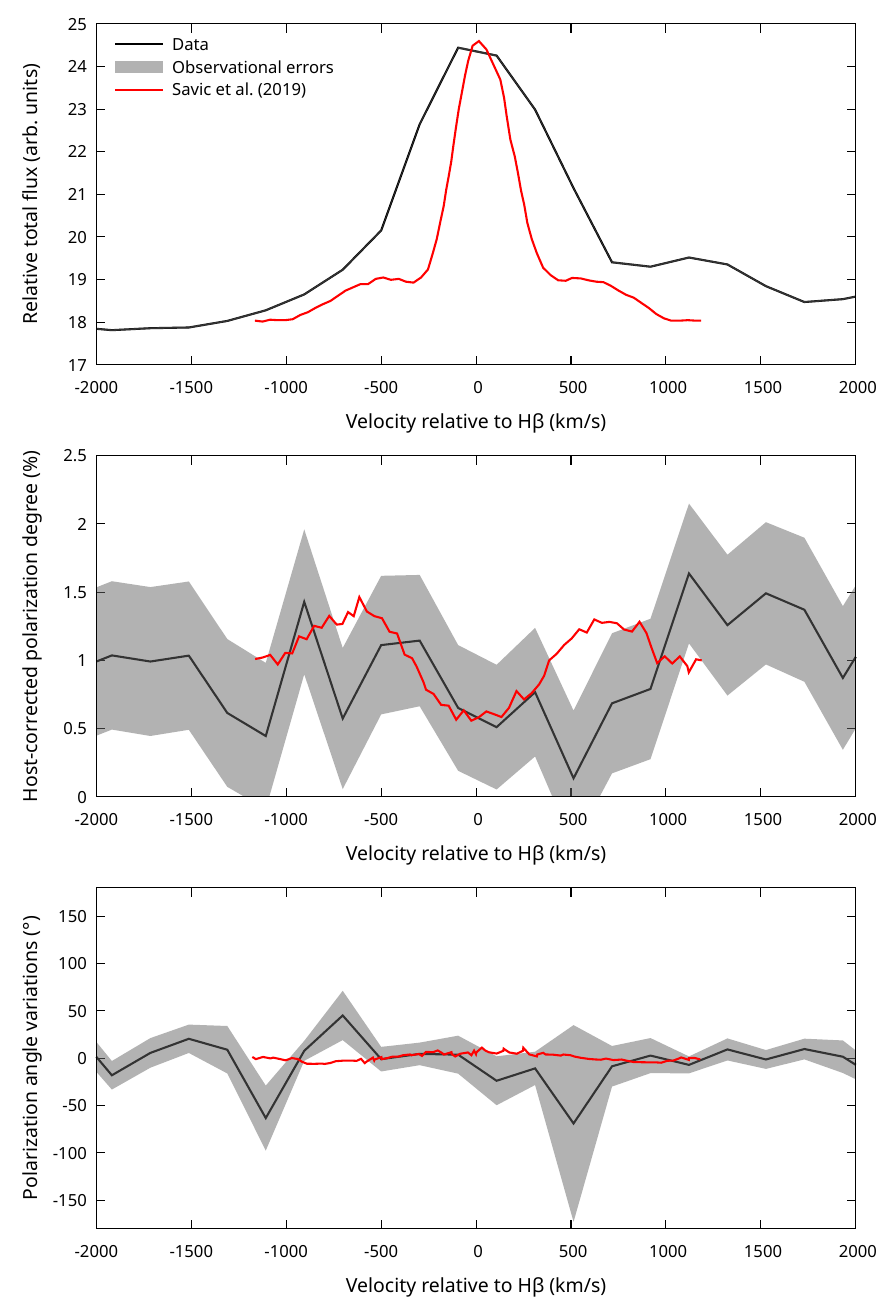}
    \caption{Observed H$\beta$ line in velocity space, indicated in black (uncertainties are shown in gray), while the predictions for a binary SMBH are overplotted in red \citep{Savic2019}. Top panel: Relative (scaled) total flux between the observation and the simulation. Middle panel: Host-corrected polarization degree of NGC~1566 and the simulation. Bottom panel: Variations of the polarization angle around 0$^\circ$ between the model and the observational data.}
    \label{Fig:appendix_Hbeta}
\end{figure*}

Figure~\ref{Fig:appendix_Hbeta} shows the H$\beta$ emission line in velocity space. The observational data are shown in black (uncertainties in gray) and the simulation in red. While the total flux profile of NGC~1566 is broadly consistent with the expected line profile from the model, the observed, host-corrected polarization degree and polarization angle remain too noisy to reveal any clear velocity-dependent structure across the line. The intrinsic polarization level predicted by the simulations is comparable to the polarization amplitude observed in our data, but the uncertainties are too large to distinguish any specific pattern in the polarization parameters. Taken in isolation, this analysis neither proves nor strictly rules out the binary SMBH hypothesis.

~\

We must remark that this represents the best data-to-simulation fit from \citet{Savic2019}. The vast majority of models predict velocity-dependent, phase-dependent polarization degrees that exceed 1\%, double-peaked, asymmetric emission line profiles and variation in $\theta$ that can reach 40$^\circ$, some features that should have appeared in our spectra despite their modest spectral resolution and S/N. In addition, when considered together with the other observational constraints discussed in the manuscript, the binary SMBH scenario appears disfavored.

\end{appendix}

\end{document}